\documentclass[12pt]{iopart}

\usepackage{graphicx}
\usepackage{cite}

\newcount\mstylenum 
\def\varstyle#1{\mathchoice{\mstylenum=0 #1}{\mstylenum=1 #1}{\mstylenum=2 #1}{\mstylenum=3 #1}}
\def\usestyle{\ifcase\mstylenum \displaystyle\or\textstyle\or\scriptstyle\or\scriptscriptstyle\fi}

\def\gensim#1{\mathrel{\varstyle{\vtop{\offinterlineskip
   \halign{\hfil$\usestyle##$\hfil\cr#1\cr\noalign{\kern1pt}\sim\cr}}}}}
\def\lesssim{\gensim<}  \def\gtrsim{\gensim>}

\begin{document}

%\textrm{\textsf{\textrm{}}}
%%%documentclass[manuscript]{revtex4}
%\documentstyle[osa,twocolumn]{revtex}
%newcommand{\MF}{{\large{\manual META}\-{\manual FONT}}}
%%%\newcommand{\manual}{rm}       
%%%\newcommand\bs{\char '134 }    
% 
%%%\usepackage{graphicx}  
%%%\begin{document}
%

\title[Theoretical analysis  
of  injection driven thermal light  emitters]{Theoretical analysis   
of injection driven thermal light  emitters based on graphene encapsulated by hexagonal boron nitride
}
\author{ 
 \bf V  Ryzhii$^{1,2}$, T  Otsuji$^1$,  M Ryzhii$^3$, V Leiman$^4$, P P Maltsev$^2$,\\
V E Karasik$^5$, V Mitin$^{6}$,  and   M~S~Shur$^7$    
 }
\address{$^1$Research Institute of Electrical Communication, Tohoku University,\\  Sendai 980-8577, Japan} 
\address{$^2$Mokerov Institute of Ultra High Frequency Semiconductor Electronics, RAS,\\ Moscow 117105, Russia}
\address{$^3$Department of Computer Science and Engineering, University of Aizu,
Aizu-Wakamatsu 965-8580, Japan}
\address{$^4$Center for Photonics and Two-Dimensional Materials, Moscow Institute of Physics and Technology, Dolgoprudny 141700, Russia}
\address{$^5$Center of Photonics and Infrared Engineering, Bauman Moscow State Technical University, Moscow 111005, Russia}
\address{$^6$Department of Electrical Engineering, University at Buffalo SUNY,\\ Buffalo, NY 14260-1920, USA}
\address{$^7$Departments of Electrical, Computer, and Systems Engineering and Physics, Applied Physics, and Astronomy, Rensselaer Polytechnic Institute,\\ Troy, NY 12180, USA}
\ead{v-ryzhii@riec.tohoku.ac.jp}

\begin{abstract} We develop the device model for the proposed injection (electrically) driven  thermal light emitters (IDLEs) based on the  vertical Hexagonal Boron Nitride Layer/Graphene Layer/ Hexagonal Boron Nitride Layer (hBNL/GL/hBNL) heterostructures and analyze their dynamic
response. The operation of the IDLEs is associated with the light emission of the hot two-dimensional electron-hole plasma (2DEHP)  generated in the GL
by both the lateral injection from the side contacts and the vertical injection through the hBNL (combined injection)
  heating the 2DEHP. The temporal variation of the  injection current results in the variation of the carrier effective temperature and their density in the GL leading to
the modulation of the output light. We determine the mechanisms limiting  the IDLE efficiency and the maximum light modulation frequency. 
A large difference between the carrier and lattice temperatures
the IDLEs with an effective heat removal enables a  fairly large modulation depth at the modulation frequencies about dozen of GHz in contrast to the standard incandescent lamps. 
We compare the IDLEs with the combined injection under consideration
and IDLEs using the carrier Joule heating by lateral current.
 The obtained results can be used for the IDLE optimization.
\end{abstract}

%\submitto{\SST}
\maketitle

%%%%%%%%%
\ioptwocol
%%%%%%%%%%
\section{Introduction} 
Heterostructures with graphene layers (GLs)
are promising building blocks for infrared and terahertz photodetectors\cite{1,2,3,4,5,6,7,8,9,10,11,12,13,14}, optical modulators~\cite{15,16,17,18},  plasmonic  and frequency multiplication devices~\cite{19,20,21,22,23,24,25,26,27,28},
 and  lasers and light-emitting diodes~\cite{29,30,31,32,33,34,35,36,37,38,39,40,41,42,43,44} (including those based on hybrid GL/black phosphorous devices~\cite{45,46}).
The realization of the on-chip monolithic nanoscale   relatively simple light sources for high-bandwidth inter- and intra-chip connections is still a challenging problem~\cite{47}.
There are several proposals and realizations of the compact and simple GL thermal light sources  using
the electrical carrier heating  in the GLs~\cite{48,49,50,51,52,53,54,55}. 
Such GL-base thermal light sources can be fairly effective and
very fast~\cite{54}. The operation and, in particular, the operation speed of the GL-based thermal sources are determined by complex relaxation and recombination mechanisms, namely,
of the carrier-carrier scattering, the interaction of the hot carriers in the GL with the GL optical phonons
and the interface optical phonons and the optical phonons in the media surrounding the GL. 
The carrier heating in the GLs  in the devices in question can be associated with the Joule heating by
the electric current flowing in the GL between the side contacts to the GL~\cite{48,49,50,51,52,53,54,55}. 
Another option
is to use the combined carrier injection (the injection of relatively cold carriers from the side contact(s)
and the vertical injection of the not too hot carriers from the top contact). 
This
concept was recently applied to the GL heterostructures
~\cite{44,45,46} aiming to realize
the interband population inversion in the GLs
for its use for far-infrared and terahertz lasing or superluminescence (the combined carrier injection is widely used in the standard vertical-cavity surface emitting heterostructure lasers).
In such heterostructure devices the vertical injection should not lead to a marked heating of the two-dimensional electron-hole plasma (2DEHP) in the GL. 
This is why the
black phosphorus (BP) (or similar materials) having relatively
small band off-sets at the heterointerface~\cite{56,57} were chosen
for the emitter contact layer.
In contrast, the thermal electrically driven
light sources (IDLEs) require strong heating of
the 2DEHP in the GLs. The thermal IDLEs, in
particularly those based on the GLs
encapsulated by the hBNLs, exhibit the
following features. First of all, both the optical
phonon and acoustic phonon systems can be
relatively hot (their effective temperatures
substantially exceed the temperature of the
thermostat) when the carrires are hot. Second,
the Auger recombination-generation processes
[58] can be much more effective than those at
moderate or low carrier temperatures.

This is why, the use of the emitter contact layer made of the black phosphorus (BP)
(or similar materials) having relatively small band off-sets at the GL/black phosphorus~\cite{56,57} interface was proposed.
 However, in contrast, in the injection driven thermal light sources (IDLEs) a strong heating of the 2DEHP in the GLs  is desirable. These thermal light sources, in particular those based  on the GLs encapsulated by the IDLEs, exhibit the following features.
First of all, at sufficiently hot carrier, both the optical phonon and acoustic phonon systems can also be relatively hot (their effective temperatures substantially exceed the temperature of the thermostat). Second,   
the  Auger recombination-generation processes~\cite{58} can be much more effective than
at moderate or low carrier temperatures.

In this paper, we study the thermal IDLEs based on the GL/hBN heterostructures with the combined
lateral/vertical carrier injection providing an effective heating due to rather large the GL/hBN band off-sets. We focus on the dynamic properties of these sources, which can be used
in different opto-electron systems. In particular, we compare the characteristics of the IDLE in question with the similar IDLE using the vertical double injection and the Joule heating and demonstrate that the IDLEs with the combined and double vertical injection can be faster due to a higher sensitivity to the variations of the controlling voltage.

The paper is organized as follows. In Section 2, we present the proposed IDLE device structure and formulate the pertinent  mathematical model. In Section 3, we use this model for the calculations of the steady-state effective temperatures of the  carrier, optical phonon, and lattice (acoustic phonon) systems as functions of the injected carrier current and the structural parameters. Apart from the numerical solution of the equations of the model, we analytically  analyze their asymptotic behavior in the limiting cases. Section~4 deals with the calculations of the  spectrum of the light emitted by the IDLEs  and the output power at the steady-state carrier injection. For this, we use
the data for the effective temperatures obtained in Section 3.
In Section 5, we analyze the  modulation of the output light by the ac injection current characteristics and evaluate
the IDLE maximum modulation frequency. Section~6 is devoted to the derivation of the light modulation depth as a function of the modulation ac voltage.
In Sections~ 7 and 8, we comment the obtained results and formulate the conclusions. Some mathematical details are singled out to the Appendix.

\section{Device structures and mathematical model}
 
\begin{figure*}[t]
\begin{center}
\includegraphics[width=12.0cm]{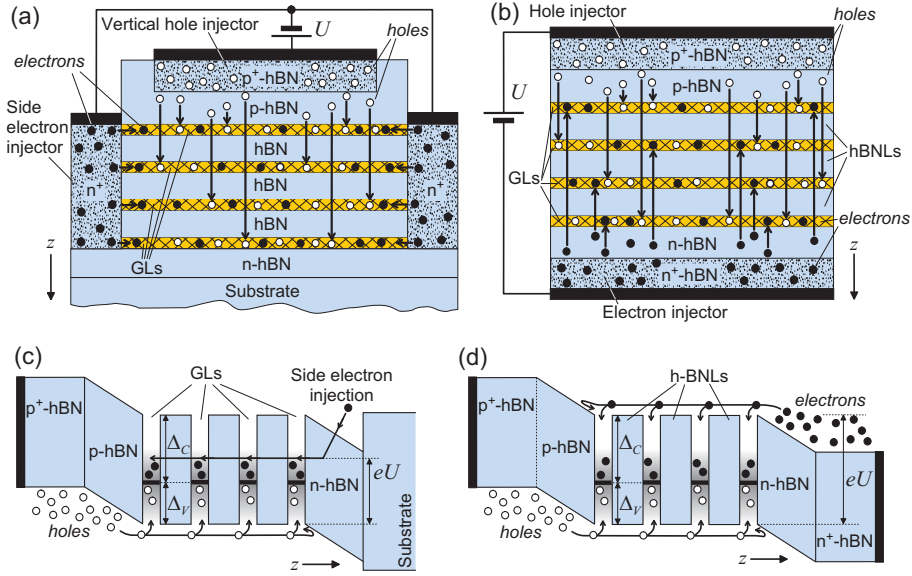}
\end{center}
\caption{Schematic views of  the IDLE based on heterostructures
(a) with the top p$^+$-p-hBN vertical hole injector and the GLs separated by the hBNLs and supplied with the n$^+$ side contacts and(b)  with the top p$^+$-p-hBN  hole  and bottom  n$^+$-n-hBN  vertical injectors 
and the band diagrams of (c) the IDLE with vertical-lateral injection and (d) vertical double injection
under applied voltage $U$. Opaque and open circles correspond to electrons and holes, respectively.
Arrows show the carrier movement directions
}
\end{figure*}

Figure 1 shows the schematic views of  the IDLEs based on the hBNL/GL/hBNL heterostructures
with the combined (vertical-lateral) and double vertical injection. 
The GLs  in the IDLE with the combined injection  are supplied by the side contacts providing the electron injection to the GLs.
The heterostructure top  comprises an undoped or lightly doped hBN layer and a  cap, which is  heavily doped  by acceptors.
Such a  p$^+$-p- hBNL region serves as an injector of holes into the GL.
Here we focus on  the IDLE with the vertical hole injection and the lateral electron injection [see figures~1(a)
and 1(c)]. The obtained results can be easily extended for the IDLEs with the combined injection can be based on the heterostructures n$^+$-n-hBN vertical
injector and the lateral hole injection and the IDLEs with the vertical double injection (see below). 
The IDLEs under consideration can comprise the multiple- and single GLs.
The device structure can be placed on a SiO$_2$ substrate.

Our mathematical model (which is the generalized version  of those
 used previously~\cite{44,45,46}, see also more early works~\cite{31,59})
corresponds to the following scenario:\\
 (i) The injected hot holes transfer their energy partially to different optical phonon modes (intra-valley and inter-valley phonons in the GL and the interface phonons.
  (emitting the cascades of these phonons) and partially to the 2DEHP via the injected hole collisions with the 2DEHP electrons and holes;\\
  (ii) The 2DEHP heated by the injected holes transfer the excessive energy to the optical  phonon systems~\cite{60};\\
(iii) The   optical  phonon system transfers the energy to the acoustic phonons (lattice) due
to the anharmonic optical and interface phonons decay~\cite{61,62,63,64,65};\\
(iv) The hot acoustic phonons bring their thermal energy primarily~\cite{55} to the top or bottom contact serving as the heat sink;\\
(v) The deviation of the  electron and hole densities from the equilibrium ones is controled by the
interband recombination-generation processes involving the optical  phonons~\cite{31,59,60} and  the Auger processes~\cite{58,66}.
 
It  is assumed that the carrier-carrier interactions in the 2DEHP are sufficiently strong, so that both the electron and hole components  have the common effective temperature $T$, which 
is generally different from the effective optical  phonon temperature $\Theta$, the acoustic phonon temperature $T_L$ [the lattice temperature around the GL(s)], and the sink (vertical contact, thermostat) temperature $T_{C}$.
Due to the latter, the electron and hole energy distribution functions are $f_e(\varepsilon_e) = \{1 + \exp[(\varepsilon_e - \mu_e)]/T\}^{-1}$
and $f_h(\varepsilon_h) = \{1 + \exp[(\varepsilon_h - \mu_h)]/T\}^{-1}$ with $\varepsilon_e$,  $\varepsilon_h$, $\mu_e$, and $\mu_h$ being the pertinent kinetic carrier energies and quasi-Fermi energies counted from the GL Dirac point.
This assumption is particularly reasonable for the IDLEs due to their operation at elevated carrier densities.
For simplicity, we do not distinguish different optical modes assuming that the optical phonon system is characterized by a single effective phonon energy $\hbar\omega_O$,
which accounts for the intra-valley (with the energy $\sim$~200~meV), inter-valley ($\sim$~165~meV), and  inteface ($\sim$~100~meV) optical phonons.  

Considering the abovementioned processes, namely the intraband and interband emission and absorption of optical phonons~\cite{62},
the decay of these phonons into the acoustic phonons,  and the Auger recombination-generation processes as the main mechanisms
determining the carrier density and energy balances in the 2DEHP,
 the main equations of the model can be presented in the following form:   
 
\begin{eqnarray}\label{eq1}
\frac{d \Sigma}{dt} = \frac{j}{e}- R_{0}^{inter}  - R_{A} ,
\end{eqnarray}

\begin{eqnarray}\label{eq2}
\frac{d{\cal E}}{dt} = \frac{j}{e}{\overline \Delta_i} - \hbar\omega_0(R_{0}^{inter} + 
R_{0}^{intra}).
\end{eqnarray}
Here $\Sigma$ and ${\cal E}$ are the 2DEHP density (the net density of the electrons and holes in the GL) and  their energy, respectively, $j$ is the density of the vertically injected current into a GL 
(in the cases of the IDLEs with a single GL and multiple GLs, $j =j_i$
and  $j = j_i/n$, respectively, where $j_i$ is the density of the net injected current and $n$ is the number
of the GLs~\cite{44}), $\overline \Delta_i$
is the energy transferred by one vertically injected holes directly to the 2DEHP (taking into account that
a fraction of the injected hole energy $\Delta_i$ goes to the optical phonons), end
$e =|e|$ is the electron charge.
The terms $R_{0}^{inter}$,  $R_{0}^{intra}$,  and $R_A$ are the rates of the 
 interband and intraband transitions associated with the pertinent optical phonons and the Auger recombination-generation processes; their explicit forms are  presented in the Appendix. 
The average energy brought to the 2DEHP by an injected hole is estimated as~\cite{58}
 
\begin{eqnarray}\label{eq3}
\overline{\Delta_i} = \Delta_i - k_0\hbar\omega_0.
\end{eqnarray}
Here $\Delta_i =\Delta_V + 3T_{0}/2$, 
 $T_0$ is the hole temperature in the p$^+$ injector, coinciding with   the metal contact temperature (which serves as the heat sink).
 Equation~(3) shows
that each injected hole energy
(with the average injected kinetic energy $\varepsilon_i = \Delta_V + 3T_{0}/2$) transferred directly to the 2DEHP is smaller than $\varepsilon_i$ because a portion, $k_0\hbar\omega_0$, of this energy goes immediately to the optical phonons. 
If we disregard the energy dependence of the time, $\tau_0$, of the optical phonon emission,  the average numbers of the optical phonons of the given sort generated by the injected hole is equal to $k_{0} =  \displaystyle\frac{1}{\tau_0/\tau_{cc}}\biggl[1 - \frac{1}{(1 +\tau_0/\tau_{cc} )^{K_{0}}}\biggr]$, where $K_0$ is the maximum numbers of the optical phonons in the cascade emitted by the hole after its injection (see the Appendix). Since the density of states in  GLs virtually linearly increases with the carrier energy, one might assume that the optical phonon intraband emission decreases with the energy. In this case,
$k_{0} = K_{0}/(1 + \tau_0/\tau_{cc})$.
Setting $\Delta_V = 1200$~meV, $T_{0} = 25$~meV, $K_0 = 8$,  one obtains
$\Delta_i \simeq  1240$ meV and
$\overline{\Delta_i} \simeq  1090 - 1210$ meV (i.e., for the above two cases, $\Delta_i \lesssim \Delta_V$). 

Considering the deviation optical  phonons distribution function ${\cal N}_{0} = [\exp(\hbar\omega_{0}/\Theta) - 1]^{-1}$  from the distribution function of optical
phonons,${\cal N}_{0}^{eq} = [\exp(\hbar\omega_{0}/T_L) - 1]^{-1}$,
in  equilibrium with the acoustic phonons in the vicinity of the GL
 (where $T_L$ is the lattice temperature or the effective temperature of acoustic phonons in the vicinity of the GL),  for  ${\cal N}_{0}$ we obtain

\begin{eqnarray}\label{eq4}
\frac{d{\cal N}_{0}}{dt}= \frac{T_L}{T_0}\frac{({\cal N}_{0}^{eq} -  {\cal N}_{0})}{\tau_D}\nonumber\\
 +
\frac{1}{\overline{\Sigma_0}}\biggl(R_{0}^{inter} + R_{0}^{intra}
+ \frac{jk_{0}}{e}\biggr).
\end{eqnarray}
Here $\overline{\Sigma_{0}}  =  (1/2\pi)(\hbar\omega_0/\hbar\,v_W)^2$ is the  characteristic carrier density determined by the energy dependence of the density of state in the GL (see the Appendix), 
$\hbar$ is the Planck  constant, and $v_W \simeq 10^8$~cm/s  is the electron and hole velocity in GLs. The quantity $\overline \Sigma_0/\tau_0$ is estimated using the data on the carrier interband generation-recombination rate~\cite{59} and $\tau_D$ is the time of the optical phonon decay due to the unharmonic lattice processes at $T_L = T_0$.
 Theoretical and experimental studies for $\tau_D$ in GLs at room lattice temperature~\cite{60,61,62,63} yield the values in the range from 1 to 5~ps. 
However,
this time can be much shorter: $\tau_D \simeq 0.20-0.35$~ps
in the GLs encapsulated by the hBNLs due to
the role of the interface optical phonons. 
The factor $T_L/T_0$ reflects an increase in the optical phonon decay rate (an increase in the decay time) with rising lattice temperature (the effective temperature of acoustic phonons)~\cite{67}. 
This temperature dependence arises from the
dependence of the unharmonic decay process
on the "daughter" acoustic phonon modes pop-
ulation~\cite{67,68}.
The lattice temperature TL is determined
by the power injected into the GL and
by the thermal conductivity the per unit
area, C, of the layers surrounding the GL,
i.e., the hBNLs and the metal contact.
Neglecting the lateral heat transfer to the side contacts~\cite{48,49,50},
the lattice temperature, $T_L$, around the GL can be determined from the following equation:

\begin{eqnarray}\label{eq5}
C\tau_L \frac{dT_L}{dt}= C(T_{0} - T_L) +  \frac{j\Delta_i}{e}.
\end{eqnarray}
Here $\tau_L$ is the characteristic time of the lattice
heating and cooling down that is proportional to
the lattice heat capacitance $c_L$.

The efficiency of the 2DEHP, optical phonon systems, and the lattice temperature crucially depends on the thermal conductivity $C$, i.e., on the the hBNLs thermal conductance of $c$ ($C \propto c$). 
If  the metal contact really serves as the heat sink, i.e., if the contact temperature is close to room temperature, considering  the hBNLs thermal conductance of $c \simeq 20$W/m$\cdot$ K~\cite{69},
for the hBNLs of the thickness $L_{hBN} = (1 - 2)~\mu$m, we obtain $C =c/L_{hBN} \simeq (1 - 2) $~kW/cm$^2$K.
However, when the cooling down of the top  metal contact is limited by the heat transfer to the surrounding air, the quantity $C $ can be several orders of magnitude smaller.

\section{Steady-state effective temperatures}

\subsection{Numerical analysis}
\begin{figure}%Fig.2
\begin{center}
\includegraphics[width=8.0cm]{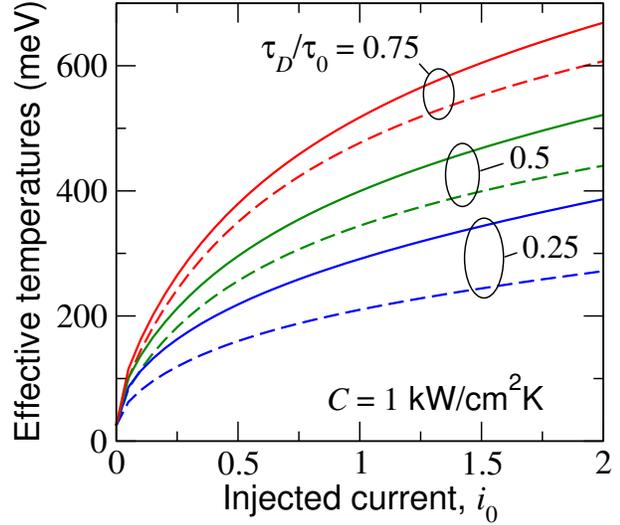}\\
\end{center}
\caption{Effective   carrier temperature $T$ (solid lines) and optical phonon temperature $\Theta$ (dashed lines) versus the normalized injected current $i_0$
for $\Delta_i = 1240$~meV,  $\overline{\Delta_i} = 1210$~meV, $C = 1$~kW/cm$^2$K, $T_0 = 25$~meV,
and  different values of the optical phonon decay time $\tau_D/\tau_0 = 0.25 -0.75$~ps. 
}
\end{figure}

Consider first the case of the dc injection current  $j = j_0 = const$.
At $j_0 = 0$, equations~(4) - (6) naturally yield ${\cal N}_{0} = {\cal N}_{0}^{eq}$, i.e., $\Theta = T_L =T_{0}$,
and $\mu_e + \mu_h = 0$. 
Using the steady-stater versions of  equations~(1) and (2) (see the Appendix), we obtain
 
\begin{eqnarray}\label{eq6}
 \biggl[({\cal N}_{0} + 1)\exp\biggl(\frac{\mu_e + \mu_h -\hbar\omega_{0}}{T} \biggl) -{\cal N}_{0} \biggr]
\nonumber\\
+\frac{\tau_0}{\tau_A}\biggl[\exp\biggl(\frac{\mu_e + \mu_h}{T}\biggr) - 1\biggr] = i_0,
\end{eqnarray}

\begin{eqnarray}\label{eq7} 
 \biggl[({\cal N}_{0} + 1)\exp\biggl(\frac{\mu_e+\mu_h}{T}\biggr)\exp\biggl(-\frac{\hbar\omega_{0}}{T} \biggl) -{\cal N}_{0} \biggr]\nonumber\\
+
\frac{1}{\eta_0} \biggl[({\cal N}_{0} + 1)\exp\biggl(-\frac{\hbar\omega_{0}}{T} \biggl) -{\cal N}_{0} \biggr]
 = i_0\frac{\overline{\Delta_i}}{\hbar\omega_0}. 
\end{eqnarray}
Here $i_0 = j_0/\overline{j}$, where  $\overline{j} = e\overline{\Sigma_0}/\tau_0 = eG_0\exp{(\hbar\omega_0/T)}$,
$\eta_{0} \simeq (\hbar\omega_{0}/\pi\,T)^2$~\cite{59}.
The quantity $\eta_{0}$ characterizes the relative contributions to the 2DEHP energy relaxation
due to the  interband and intraband transitions mediated  by the optical phonons. In the case of rather hot carriers, the intraband processes  involve wide energy ranges, so that $\eta_{0}$ can be close to unity. The characteristic injection current density $\overline{j}$, determined by the parameters of carrier scattering on different optical phonon modes, can be treated as a main fitting parameter of the model under consideration.
For definiteness, in the following assuming that the rate of the electron-hole pairs
interband generation due to the absorption of the thermal optical phonons at $T_0 = 25$ ~meV and the characteristic ("average") photon energy are equal to $G_0 = 2\times 10^{21}$~cm$^{-2}$s$^{-1}$ and  $\hbar\omega_0 = 150$~meV, respectively (see the Appendix). We also set $\overline{\Sigma_0}/\tau_0 = 2\times 10^{24}$~cm$^{-2}$s$^{-1}$,
so that $\tau_0 \simeq 0.42$~ps and $\overline j \simeq  320$~kA/cm$^{2}$.

Due to a large ratio $\Delta_i/\hbar\omega_0$ and  a smallness
of $\tau_A/\tau_0$ in the situations under consideration, the right-hand side of equation (A6) can be simplified and we obtain, so that from equations~(6) and (7)

\begin{eqnarray}\label{eq8} 
\biggl(1 +\frac{1}{\eta_0}\biggr) \biggl[({\cal N}_{0} + 1)\exp\biggl(-\frac{\hbar\omega_{0}}{T} \biggl) -{\cal N}_{0} \biggr]\nonumber\\
 \simeq i_0\biggl(\frac{\overline{\Delta_i}}{\hbar\omega_0}
\biggr). 
\end{eqnarray}

Accounting for the explicit expression for the terms in its right-hand side 
(see  the Appendix) equation~(4) yields

\begin{eqnarray}\label{eq9}
{\cal N}_{0}= \frac{1}{\exp(\hbar\omega_0/T_L) - 1} 
+i_0\frac{\Delta_i}{\hbar\omega_0}\frac{\tau_D}{\tau_0}\frac{T_0}{T_L},
\end{eqnarray}
or

\begin{eqnarray}\label{eq10}
\Theta = \frac{\hbar\omega_0}
{\ln\Biggl\{1 +\frac{\displaystyle\exp\biggl(\frac{\hbar\omega_0}{T_L}\biggr) - 1}{1 + \displaystyle i_0\frac{\Delta_i}{\hbar\omega_0}\frac{\tau_D}{\tau_0}\frac{T_0}{T_L}\biggl[\exp\biggl(\frac{\hbar\omega_0}{T_L}\biggr) - 1\biggr]}
\Biggr\}}.
\end{eqnarray}
Here $T_L$ obeys  the following equation [see equation~(5)]:

\begin{eqnarray}\label{eq11}
T_L = T_0\biggl[1 + \beta\,i_0\biggl(\frac{\Delta_i}{T_0}\biggr)\biggr],
\end{eqnarray} 
where $\beta = \overline \Sigma_0/\tau_0\,C$.
In the case of an 
undoped GL, $\mu_e = \mu_h$, so that for $2\mu = \mu_e + \mu_h$
one can obtain
[with the same accuracy in deriving equation (9)]:
 
\begin{eqnarray}\label{eq12} 
\exp\biggl(\frac{2\mu}{T}\biggr) \simeq 1 + i_0\frac{\tau_A}{\tau_0}.
\end{eqnarray}

Equations~(8) - (11)  
allow to find the the effective carrier and optical phonon temperatures, $T$ and $\Theta$, as well as the lattice temperature $T_L$ near the GL  
 as functions of the dc  injected current density  for the IDLEs with different structural parameters.

%%%%%%%%%%%%%%%%%%%%%%%%%%%%%%%%%%%%%%%%%%%%%%%

%%%%%%%%%%%%%%%%%%%%%%%%%%%%%%%%%%%%%%%%%%%%%%

Figure~2 shows the dependences of the effective temperatures of the carriers $T$,
 optical phonons, and lattice $T_L$ 
 $\Theta$ 
 on the normalized injected  current density $i_0$ (i.e., actually on the injected power density $\Delta_i\,j_0/e$)
  for  different values of $\tau_D/\tau_0$ calculated using equations~(8) - (12).  
It is assumed that $T_{0} = 25$~meV, $\hbar\omega_0 = 150$~meV, 
$\Delta_i = 1240$~meV,
$\overline{\Delta_i} = 1210$~meV,
%$\tau_0 = 1$~ps, 
 $C = 1$~kW/cm$^2$K, and $\beta \simeq 2.76\times 10^{-2}$.

Figure~3 shows  $T$, $\Theta$, and $T_L$ 
versus the normalized injected  current density $i_0$ for   $\Delta_i = 1240$~meV,   $\overline {\Delta_i} =  1090$ and  1210~meV (corresponding to $\tau_0/\tau_{cc} = 1 - 5$),
 assuming  $C = 1$~kW/cm$^2$K.
One can see that $T$ at $\overline{\Delta_i} = 1210$~meV only slightly exceeds
$T$ at $\overline{\Delta_i} = 1090$~meV, while the values of  $\Theta$ for
these cases are indistinguishable. As seen from figure~3, the lattice temperature $T_L$ is much lower than $T$ and $\Theta$ (for the chosen value of $C$).
The lattice temperature around of the GL  is independent of $\tau_D$. As follows from equation~(11) (and figure~
3), it increases from $T_L = 25$~meV 
(300 K) at $i_0 = 0$  to $T_L \simeq 59$~meV (about 710 K) at $i_0 =1$, and                                        to $T_L \simeq 93$~meV (about 1116 K) at $i_0 =2$.
 Hence, in the injected current densities  range under consideration and the relatively high $C$ characteristic for the hBNLs, $T_L \simeq < \hbar\omega_0$.
However, at small  values of the hBNL thermal conductivity $C$ (in the devices with far from ideal heat removal), $T_L$ can be very close to $\Theta$ and only slightly lower than $T$. 
Figure~4 shows  $T$, $\Theta$, and $T_L$
calculated as functions of the hBNL thermal conductivity (i.e., the hBNL thickness).
One can see that in a wide range of $C$, the temperatures $T$, $\Theta$, and $T_L$ fairly weakly depend on $C$ in a wide range of its variation (at the given
injection current densities). 
 However, at small values of the hBNL thermal conductivity $C$(the devices with 
 far from ideal thermal removal), 
the temperatures $T$, $\Theta$, and $T_L$
can become quite  high with 
 $T_L$ being very close to $\Theta$ and only slightly lower than $T$.
The $T$, $\Theta$, and $T_L$ versus $C$ relations are shown only in the range $C$
($C \geq 0.10 - 0.17$~kW/cm$^2$K) corresponding to $T_L$ smaller than
 the hBNL melting temperature  $T_L \lesssim  250$~meV ($T_L < 2973$~K).

\begin{figure}%Fig.3
\begin{center}
\includegraphics[width=8.0cm]{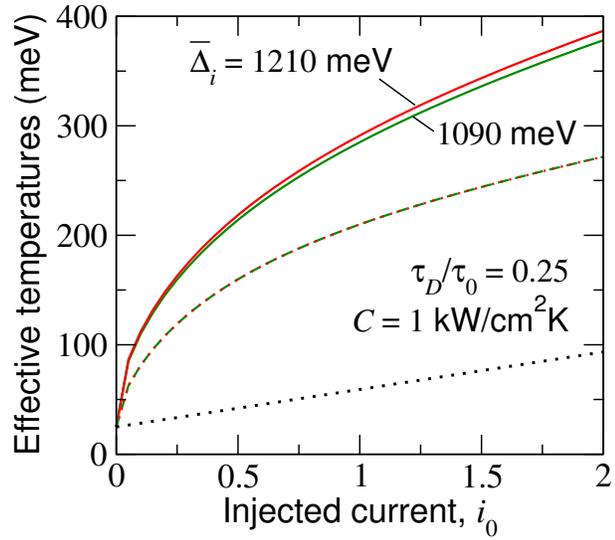}
\end{center}
\caption{Effective   carrier temperature $T$ (solid lines), optical phonon temperature $\Theta$ (dashed line), and lattice temperature $T_L$ (dotted line)
versus the injected current $j_0$ for  $\tau_D/\tau_0 = 0.25$  at
   different    $\overline{\Delta_i}$, setting $\Delta_i = 1240$~meV,
 $C = 1$~kW/cm$^2$K, and $T_0 = 25$~meV.
}
\end{figure}

\begin{figure}%Fig.4
\begin{center}
\includegraphics[width=8.0cm]{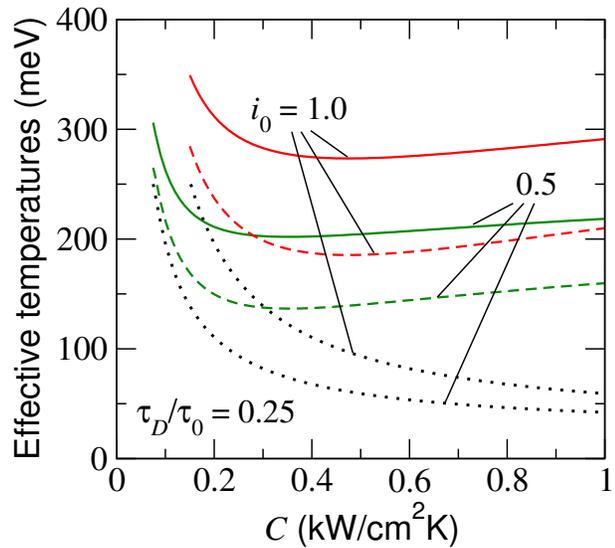}
\end{center}
\caption{Effective   carrier temperature $T$ (solid lines), optical phonon temperature $\Theta$ (dashed lines), and  lattice temperature $T_L$ (dotted lines)  as functions of the hBNL thermal conductivity $C$ at the  injected current  $i_0 = 1$ and  $i_0 = 2$
for $\Delta= 1240$~meV, $\overline{\Delta_i}= 1210$~meV, $\tau_D/\tau_0 = 0.25$, and $T_0 = 25$~meV.}
\end{figure}

\subsection{Asymptotic analytic relations: weak heating}

One can  get analytic formulae for $T$ and $\Theta$ in the limits of small and relatively high injection currents.
At small injected current densities $j_0$ ($i_0 \ll 1$), the temperatures  $T, \Theta, T_L \ll  \hbar\omega_0$ and
 equation~(10) yields

\begin{equation}\label{eq13}
\Theta \simeq T_0\biggl[1  + i_0\frac{\Delta_i}{\hbar\omega_0}\frac{T_0}{\hbar\omega_0}\exp\biggl(\frac{\hbar\omega_0}{T_0}\biggr)\frac{\tau_D}{\tau_0}\biggr].
\end{equation}

As for the carrier effective temperature $T$ at relatively low injection current,
from equation~(8) we obtain 

\begin{eqnarray}\label{eq14}
T \simeq  \Theta + T_0\frac{\displaystyle i_0\frac{\overline \Delta_i}{\hbar\omega_0}\biggl(\frac{T_0}{\hbar\omega_0}\biggr)\exp\biggl(\frac{\hbar\omega_0}{T_0}\biggr)}{\biggl[1  + \displaystyle\biggl(\frac{\pi\,T_0}{\hbar\omega_0}\biggr)^2\biggr]}
\end{eqnarray}
with $\Theta$ given by equation~(13).
According to equation~(14), $T > \Theta$. 
An increase in $T$ with increasing ijection current density is consistent
with the energy relaxation on 
 optical phonons  in line with the previous results~\cite{18,70}.

\subsection{Asymptotic analytic relations: strong heating}

At a strong injection (large $i_0$), the 2DEP, optical phonon system and the lattice (near the GL) can be fairly hot. At large values of $i_0$, from equation  (8) - (12) we obtain the following relations:

\noindent(A) High  thermal conductivity $C$ (small~$\beta$). In this case, 

\begin{eqnarray}\label{eq15}
T_L \simeq T_0,\qquad \Theta \simeq i_0\Delta_i\frac{\tau_D}{\tau_0},
\end{eqnarray}
\begin{eqnarray}\label{eq16}
T \simeq \frac{i_0\Delta_i}{2}\biggl(\frac{\tau_D}{\tau_0}\biggr)
%\nonumber\\
%\times
\biggl[1 + \sqrt{1 + \frac{4}{\pi^2}\frac{\overline \Delta_i}{\Delta_i}\frac{\hbar\omega_0}{\Delta_i}
\biggl(\frac{\tau_0}{\tau_D}\biggr)^2 \frac{1}{i_0}}\biggr]\nonumber\\
\simeq i_0\Delta_i\biggl(\frac{\tau_D}{\tau_0}\biggr) +\frac{\hbar\omega_0}{\pi^2}\biggl(\frac{\tau_0}{\tau_D}\biggr)\biggl(\frac{\overline\Delta_i}{\Delta_i}\biggr)
. 
\end{eqnarray}
%even at   fairly wide range of  $i_0$ (but $i_0 < T_0/\beta\Delta_i$).

\noindent(B) Low  thermal conductivity  $C$ (large~$\beta$). In such a case,

\begin{eqnarray}\label{eq17}
T_L \simeq \beta\,i_0\Delta_i, \qquad \Theta \simeq \beta\,i_0\Delta_i
 +
\frac{T_0}{\beta}\frac{\tau_D}{\tau_0},
\end{eqnarray}

\begin{eqnarray}\label{eq18}
T \simeq \frac{\beta\,i_0\Delta_i}{2}
%\nonumber\\
%\times
\biggl[1 + \sqrt{1 + \frac{4}{\pi^2\beta}\frac{\overline \Delta_i}{\Delta_i}\frac{\hbar\omega_0}{\Delta_i}
\biggl(\frac{\tau_0}{\tau_D}\biggr) \frac{1}{i_0}}\biggr]\nonumber\\
\simeq \beta\,i_0\Delta_i +\frac{\hbar\omega_0}{\pi^2}\biggl(\frac{\tau_0}{\tau_D}\biggr)\biggl(\frac{\overline\Delta_i}{\Delta_i}\biggr). 
\end{eqnarray}

In the  case "B", $T \gtrsim \Theta \gtrsim T_L \gg T_0$
in agreement with the  $T$, $\Theta$, and $T_L$ versus $C$ relations shown in figure~4.
Since $T_L$ is limited by the hBNL melting temperature $T_L^{melt}$,
equation~(17) provides a limitation on $i_0 < i_0^{melt} = T_L^{melt}/\beta\Delta_i$. Setting $T_L^{melt} \simeq 256$~meV ($\sim 2973$~K),  $\Delta_i = 1240$~meV,
and
$C = 0.1$~kW/cm$^2$K ($\beta = 2.76$), we find $ i_0^{melt} \simeq 0.74$.

It is instructive that according to equations~(15) - (18), both $T$ and $\Theta$ are linear functions of $i_0$ when the latter is sufficiently large.
In both cases  "A" and  "B",  the difference
$T-\Theta$ [the second terms in the right-hand sides in equations~(17) and (18)] is independent of $i_0$.
As follows from the above equations that $dT/di_0 > d\Theta/di_0$ in the range of small values of $i_0$, whereas $dT/di_0 = d\Theta/di_0$ when $i_0$ is large.
The asymptotic relations of  $T$ and $\Theta$  on $i_0$ are in line with the above numerical data.

\section{Steady-state spectral characteristic and output optical power}

The probability, $\nu_R$, of the interband radiative transition in the GL is given by~\cite{71,72}

 \begin{equation}\label{eq19}
\nu_R(p) = \frac{8}{3}\biggl(\frac{e^2\sqrt{\kappa_{hBN}}}{\hbar\,c}\biggr) \biggl(\frac{v_W}{c}\biggr)^2\frac{v_Wp}{\hbar}.
\end{equation}
Here $\kappa_S$ is the dielectric constant of  hBN, $c$ is the speed of light in vacuum, and $p =\hbar\omega/2c$ is the carrier momentum in the initial and final state, where $\hbar\omega$ is the energy of the emitted photon. In the range of the photon frequencies ($\omega > 1/\tau$)the   processes associated with the indirect
intraband transitions with the absorption (the Drude absorption) and emission of the photons are much weaker than
the interband processes. 
Therefore, we account solely for the interband radiative transitions.
Since the rate of the interband radiative transition~\cite{71,72} 

\begin{eqnarray}\label{eq20}
R_R(\hbar\omega) \propto ({\cal N}_{\hbar\omega} + 1)\nonumber\\
\times
\nu_R(p)\biggr|_{p= \hbar\omega/2c} f_e(\varepsilon_e)\biggr|_{\varepsilon_e =\hbar\omega/2} f_h(\varepsilon_h)\biggr|_{\varepsilon_h =\hbar\omega/2},
\end{eqnarray}
where ${\cal N}_{\hbar\omega} = {\exp(\hbar\omega/T_{0}) - 1}^{-1}$ is the  distribution function
of the photons in equilibrium with the thermostat,
we arrive at the following expressions for the flux  $S_{\hbar\omega}$ (in units cm$^{-2}$s$^{-1}$)
and the output power $P = A \int_0^{\infty}d(\hbar\omega) S_{\hbar\omega}$ (in units W cm$^{-2}$)
of the  photons emitted by the GL(s):

\begin{eqnarray}\label{eq21}
S_{\hbar\omega} \simeq \frac{S_0\displaystyle\biggl(\frac{\hbar\omega}{T_{0}}\biggr)^3}
{\biggl[1 + \displaystyle\zeta\exp\biggl(\frac{\hbar\omega}{2T}\biggr)\biggr]^2}
\frac{\displaystyle\exp\biggl(\frac{\hbar\omega}{T_{0}}\biggr)}
{\biggl[\displaystyle\exp\biggl(\frac{\hbar\omega}{T_{0}}\biggr) - 1\biggr]},
\end{eqnarray}

\begin{eqnarray}\label{eq22}
P = S_0T_0
\biggl(\frac{T}{T_{0}}\biggr)^4\nonumber\\
\times\int_0^{\infty}\frac{dZZ^3}{\biggl[1 + \zeta\displaystyle\exp\biggl(\frac{Z}{2}\biggr)\biggr]^2}
\frac{\displaystyle\exp\biggl(Z\frac{T}{T_{0}}\biggr)}
{\biggl[\displaystyle\exp\biggl(Z\frac{T}{T_{0}}\biggr) - 1\biggr]}.
\end{eqnarray}
Here $S_0 = \displaystyle\frac{2A}{3\pi}\biggl(\frac{e^2\sqrt{\kappa_{hBN}}\,T_{0}^3}{\hbar^4c^3}\biggr)$, $\zeta = \exp(-\mu/T)$, and $A <1$ is the fraction of the emitted photons not reflected by the outer surface. According to equation~(11),
$\zeta = (1 + \beta\,i_0\tau_A/\tau_0)^{-1/2}$. 
For $\sqrt{\kappa_{hBN}} \simeq 2.5$, $T_{0} = 25$~meV, and $A \lesssim 1$ one obtains $S_0 \simeq 3\times 10^{17}$~cm$^{-2}$s$^{-1}$.

In deriving equation~(22), we have neglected the photon recycling  and the photon accumulation inside the device (i.e., the resonant properties of the heterostructure). 
In the case of small $(\mu_e + \mu_h)/T$, equations~(21) and (22) coincide with  the pertinent formulae in our recent paper~\cite{46} [in which the injection of relatively low energy carriers was considered, so that$(\mu_e + \mu_h)/T$ can be  large].

Figure 5  shows the spectral dependence of the output radiation 
$S_{\hbar\omega}$ for different normalized injection current densities $i_0$.
Figure 6 presents the output power $P$ versus the normalized injection current density $i_0$ for the same parameters as in figure~5. 
As follows from figures~(5) and (6), the IDLEs can be effective sources of near-infrared and visible light.

The output radiation power as a function of the 2DEHP effective temperature $T$
 can be estimated by evaluating the integral in equation~(22). As a result, we arrive at

\begin{eqnarray}\label{eq23}
P \simeq
5.24\biggl(\frac{T}{T_0}\biggr)^4 {\rm mW/cm}^2.
\end{eqnarray}
The output power, $P^{BB} =
(\pi^2/ 60 \hbar^3c^2)T^4$, of the thermal radiation emitted by a similar structure  considered as 
 the black-body   with the same temperature $T$ is given by

\begin{eqnarray}\label{eq24}
P_0^{BB} = \biggl(\frac{\pi^2}{60\hbar^3c^2}\biggr)T_0^4 = 45.9~{\rm mW/cm}^2.
\end{eqnarray}
Comparing $P$ and $P^{BB}$, given by equations (23) and (24), respectively,
we obtain the following estimate for the IDLE emissivity: $\epsilon_{GL} = P/P^{BB} \simeq 0.114$.
The latter is several times larger than that estimated in ~\cite{48,49} and close to the value obtained in~\cite{46}. 

The power $P^{GB}_L$, emitted by the heated hBHL(s) considering it as a gray body with the emissivity $\epsilon_{hBN}$ and setting that the average temperature $\overline T_L
=(T_L - T_0)/2 $, where $T_L$ is given by equation~(11), is estimated as  

\begin{eqnarray}\label{eq25}
P^{GB}_L = \epsilon \biggl(\frac{\pi^2}{960\hbar^3c^2}\biggr)^4 (\beta\,i_0\Delta_i)^4.
\end{eqnarray}
For $\Delta_i = 1240$~meV, $C =  1 $~kW/cm$^2$K ($\beta = 0.0276$) and $\epsilon = 0.1$, equation~(25) yields $P^{GB}_L \simeq 1\times i_0^4$~mW/cm$^2$. Taking into account that $T > \overline T_L, T_L$, we find $P^{GB}_L \ll P$.
At low value of the thermal conductivity $C$, 
the lattice can be strongly heated, so that $\overline T_L \simeq T_L \simeq T$, and the net emitted power
can be fairly high being only limited by the melting
of the device.

The emitted optical power is much larger than that in the case of "cold" injection~\cite{46}, but the values of the emissivity are
 close in the IDLEs in question  and in the source considered in ~\cite{46},
 but the values of the emissivities of the IDLEs
and of the source considered in [46] are close.

\begin{figure}%Fig.5
\begin{center}
\includegraphics[width=7.0cm]{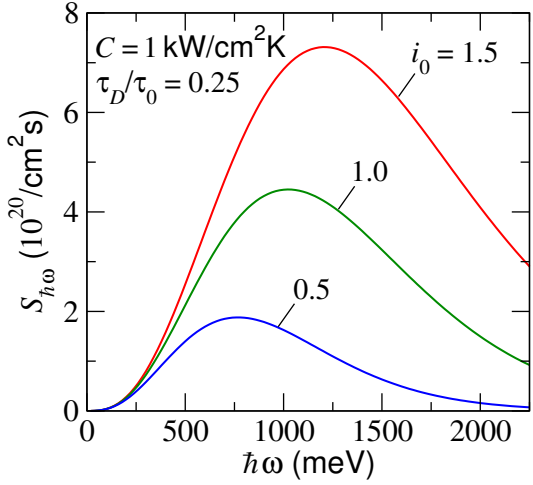}
\end{center}
\caption{Emitted radiation spectral characteristic  $S_{\hbar\omega}$ for
different normalized injection current densities $i_0$ 
($\Delta_i= 1210$~meV, $\overline{\Delta_i}= 1210$~meV, $\tau_D/\tau_0 = 0.25$, and $T_0 = 25$~meV).
}
\end{figure}

\begin{figure}%Fig.6
\begin{center}
\includegraphics[width=7.0cm]{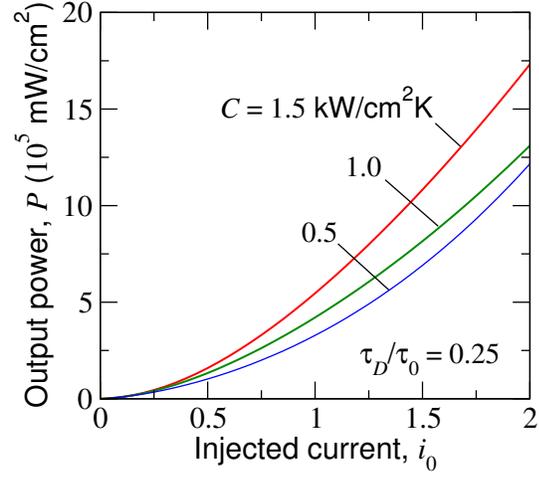}
\end{center}
\caption{Output  radiation power (per unit of square) 
as a function of 
 normalized injection current densities $i_0$ for different values of the thermal conductivity $C$.  
($\Delta_i= 1240$~meV, $\overline{\Delta_i}= 1210$~meV, $\tau_D/\tau_0 = 0.25$, and $T_0 = 25$~meV).
%Dashed line corresponds to the black-body radiation.
}
\end{figure}

\section{Dynamic response: modulation characteristics}

We consider  
 the temporal variations (modulation) of the applied voltage, $U = U_0 + \delta U_m\exp(-i\omega_m t)$,
so that $j  = j_0 +\delta j_m\exp(-i\omega_m t)$, 
 $\delta\,i_m = \delta\,j_m/\overline j$, where 
$V_0$ and $j_0$ are the dc bias voltage and current density,
$\delta V_m$ and $\delta j_m$ are the pertinent amplitudes of the ac signal, and $\omega_m$
is the modulation frequency.

In the most interesting situations when the modulation frequency $\omega_m$ is sufficiently
high [$\omega_m \gg 1/\tau_L$, see equation~(5)], the lattice temperature $T_L$ fails to follow the variations of the injected current. This implies that we can
neglect the value $\delta\,T_L$. Taking this into account, from equation (4)
we obtain
  
\begin{eqnarray}\label{eq26}
\delta {\cal N}_0 = \frac{\Delta_i}{\hbar\omega_0}\frac{\tau_D}{\tau_0}\frac{T_0}{T_L}\frac{\delta\,i_m}{(1 - i\omega_m\tau_DT_0/T_L)}.
\end{eqnarray}
If $$
\frac{\tau_D}{\tau_0}\frac{T_0}{T_L} = \frac{\tau_D}{\tau_0}\frac{1}{1 + \beta\,i_0(\Delta_i/T_0)}
\ll 1,
$$ (i.e., at
a relatively small $\tau_D$ and not too small injection current densities),
the variation of $T$ is larger than the variation of ${\cal N}_0$.
Assuming the latter for  the subsequent formulae brevity,we found from equation~(2)

\begin{eqnarray}\label{eq27}
-i\omega_m\biggl(\frac{\partial{\cal E}}{\partial T}\biggr) \delta T_m \simeq \frac{\delta j_{m}}{e}\overline{\Delta_i}\nonumber\\
- \hbar\omega_0 \biggl(\frac{\partial R_0^{inter}}{\partial T} + \frac{\partial R_0^{intra}}{\partial T}\biggr)\delta\,T_m.
\end{eqnarray}
Here the terms in the brackets are taken at the steady-state values of $T$ and ${\cal N}_0$.
Considering equation (8) and equation~(2) with $\cal E$ given by (see the Appendix)

\begin{eqnarray}\label{eq28}
{\cal E} \simeq \nu\,\Sigma_0 \biggl(\frac{T}{T_0}\biggr)^2T,
\end{eqnarray}
with the coefficient  $\nu \simeq 2.19$, when $T$ is a relatively large
($T \gtrsim \hbar\omega_0$, i.e. $T  \gtrsim 1800$~K),
we arrive at the following rough estimate for the effective carrier  temperature:

\begin{eqnarray}\label{eq29}
\frac{|\delta\,T_m|}{T} \sim    \frac{|\delta j_m|}{2j_0}\frac{1}{(\sqrt{(1 +\omega_m^2\tau_T^2)}},
\end{eqnarray}
where  $\tau_T \simeq \displaystyle\frac{\nu\tau_0}{2}\biggl(\frac{\pi\,T}{\hbar\omega_0}\biggr)^2\biggl(\frac{T}{\overline \Delta_i}\biggr)$.
At  $T = (290 - 370)$~meV ($i_0 = 1 - 2$, see figures~2 and 3) one obtains
$\tau_T \simeq 10 - 20$~ps,
For the pertinent modulation cut-off frequency $f_m^{max}$, defined as $1/\sqrt{1 +(2\pi\,f_m^{max}\tau_T)^2} = \sqrt{2}/2$, we find 
$f_m^{max}  \simeq (8 - 16)$~GHz.  The latter values of $f_m^{max}$ are in line with the
recent predictions for the GL-based light emitter with the Joule heating~\cite{55}
and much larger than those in an earlier work~\cite{73}.

Using equation~(27) for $\delta T_m$ and equation~(22) for the output optical power,
we obtain the following estimate for the radiation modulation depth $\delta P_m$:

\begin{eqnarray}\label{eq30}
\frac{\delta\,P_m}{P} \simeq  \frac{4|\delta T_m|}{T} \simeq 
\frac{2\delta j_m}{j_0}\frac{1}{\sqrt{(1 +\omega_m^2\tau_T^2)}}.
%{\rm W/cm}^2.
\end{eqnarray}

The power modulation depth described by equation~(30) is much larger that for  the devices in which the electron and lattice temperatures are close.  
The latter one 
applies to
 the IDLEs with the weak remove of the lattice heat (small $C$, large $\beta$). This is because when the steady state temperatures $T \sim T_L$, the thermal emission of the lattice (as a gray body system) makes the denominator in equation (30) too large, while $|\delta T_m| \gg |\delta T_L|$. This emphasizes the
necessity to provide a high thermal
conductance of the hBNLs and of the top metal
contact serving as effective heat sinks.
A similar situation  takes place in the standard incandescent lamps.

\section{Injected current-voltage characteristics}

The injected current density $j$ is determined by the voltage $U$ applied between
the n$^+$-side contacts to the GL(s) and the  p$^+$-region of the vertical contact (see figure 1). When $U < U_{bi}$, where $eU_{bi} \sim \Delta_G^{hBN}$ is the built-in voltage and $\Delta_G^{hBN}$ is the hBNL energy gap [as shown in figures 1(c) and 1(d)],
the hole injection into the GL(s) is controlled by the energy barrier at the p-hBNL/GL interface. High
carrier densities in the GL(s) result in high
efficiency of the carrier-carrier scattering, and,
therefore, the injected holes are more likely to be
captured into the GL(s) than rejected back to
the injector.
This is in contrast to the situations in the standard quantum-well and  GL-based~ structures~\cite{74,75} with the injected carriers mainly passing across such structures,
in which the capture probability is relatively small. 
Hence, in the IDLEs under consideration, the density of the injection current to a single- or multiple-GL is
given by 

\begin{eqnarray}\label{eq31}
j_i = j^{max} \exp\biggl[\frac{e(U - U_{bi})}{T_0}\biggr].
\end{eqnarray}
Here $j^{max} = e N^+v_T$ with $N^+$ and $v_T$ being the hole density in the p$^+$-contact region and the hole thermal velocity, respectively.
For an effective 2DEHP heating ($i_0 \gtrsim 1 - 2 $) the injected current density $j_i$
should exceed $n\overline j = e\overline\Sigma_0/\tau_0$, where $n$ is the number of the GLs in the device. Assuming the latter
be equal to  $\overline j = 320$~kA/cm$^2$ (see section 3), we arrive at the
following condition: $N^+ \gtrsim 2n\times 10^{17}$~cm$^{-2}$.
Taking into account the energy of the injected holes, we estimate
the injected power density $P_i$ for $i_0 = 0.5 - 2.0$: $P_i \simeq 200 - 800$~kW/cm$^2$. These values of the injected power density are of the same order of magnitude as, for example,  in~\cite{55}.

Equations~(30) and (31)  yield

\begin{eqnarray}\label{eq32}
\frac{\delta\,P_m}{P}\simeq \frac{2e\delta U_m}{T_0}\frac{1}{\sqrt{(1 +\omega_m^2\tau_T^2)}}.
\end{eqnarray}

\section{Comments}

Depending on the contact metal,
roughness of  of its surface  and the surrounding air pressure,  $C$ and $\beta$ can be in ranges of (1 - 15)~W/cm$^2$K and $ 2 - 30$, respectively. 
However, in the latter case, the IDLE energy balance  essentially depends
on the radiated power~\cite{50}, which is disregarded in  our model.

Similar IDLEs can exploit the lateral hole  injection combined with the vertical electron injection. In this case, the electrons captured into the GL acquire
the energy $\Delta_C$, so that in the above equations one needs to put $\Delta_i \simeq \Delta_C +3T_{0}/2$.
Due to $\Delta_C > \Delta_V$ at the hBNL/GL interface, a relatively strong 2DEHP heating can occur at markedly smaller injection current densities.
Analogously, in the IDLEs with the vertical double injection of both the electrons and holes [see Figure~1(b) and 1(d)], the quantity $\Delta_i$ can be  larger:
 $\Delta_i \simeq \Delta_C + \Delta_V+3T_{0} \gtrsim \Delta_G^{hBN}$. However,
 in this case, $\overline \Delta_i$ can be markedly smaller than  $\Delta_i$
 because of a larger factor $k_0$ in equation (3). 
 
The operation of the IDLSs under consideration is associated with the injection
of rather hot carriers. This is in contrast to the thermal light emitters using
the double injection of relatively cold carriers, which are  heated by the lateral current (via the Joule heating). The comparison of the IDLE under consideration
and the emitters using the Joule heating (for example,~\cite{54,55}) demonstrates    
comparable or better characteristics (the consumed power and the emission efficiency) of the former. 
An advantage of the IDLEs can be associated with more uniform spatial distribution of the injected power in comparison with the devices with the Joule
heating. In
the latter, the Joule power cannot be
distributed along the GL fairly uniformly
due to the essential non-linearity of the
lateral current-voltage characteristics~\cite{48,76}. Also,the  in the emitter with the Joule heating having 
a short GL, the hot carrier can quickly pass through the GL and release a substantial power in the collecting contacts.
The IDLEs can exhibit higher modulation efficiency
due to higher values of $\delta j_m/\delta U_m$. This is because in the case
of the lateral Joule heating, the injected current as a function of the applied
voltage tends to saturation~\cite{55} (see also~\cite{76}) that results in smaller $\delta j_m/\delta U_m$. 

The developed IDEL mathematical  model can also be applied for more detailed analyzing
of the thermal light sources using other method of the carrier heating,
including those with the Joule heating.

\section{Conclusions}

The transparent mathematical model of the
proposed IDLEs using the combined lateral-
vertical carrier injection into the GL(s)
encapsulated by the hBNLs allows to analyse,
evaluate, and optimize the device
characteristics.

The  device model under consideration
showed that the IDLEs can serve as  simple efficient light emitters. Due to the possibility of fast variations of the hot carrier injection and temperature by varying voltage,
these emitters can generate near-infrared and visible light  modulated in the GHz range 
 that is not possible for the standard incandescent lamps.
This reature enables IDLEs
can be used in different communication and signal processing optoelectronic systems.

\ack
One of the co-authors (VR) is grateful to Prof. Yu. G. Gurevich (Mexico) for fruitful discussion.

This work was supported by  
the Japan Society for Promotion of Science 
(KAKENHI $\#$16H06361), Japan; 
the Russian  Foundation for Basic Research
(Grants $\#$ 18-29-02089, $\#$ 18-07-01379), Russia;
RIEC Nation-Wide Cooperative Research Project $\#$H31/A01), Japan. 
The work at RPI was supported by the US Army Research Laboratory Cooperative Research Agreement (Project Monitor Dr. Meredith Reed) and by the US Office of Naval Research (Project Monitor Dr. Paul Maki), USA.

\section*{Appendix. General equations }
\setcounter{equation}{0}
\renewcommand{\theequation} {A\arabic{equation}}

\subsection*{1. Interband and intraband transitions rates}
The rates of the interband and intraband processes involving the optical phonons can be presented as:

\begin{eqnarray}\label{eq1A}
R_{0}^{inter}= \frac{\overline{\Sigma_0}}{\tau_0}\nonumber\\
\times\biggl[({\cal N}_{0} + 1)\exp\biggl(\frac{\mu_e + \mu_h - \hbar\omega_{0}}{T} \biggl) -{\cal N}_{0} \biggr],
\end{eqnarray}

\begin{eqnarray}\label{eqA2}
R_{0}^{intra}= \frac{\overline{\Sigma_0}}{\tau_0\eta_{0}}\biggl[({\cal N}_{0} + 1)\exp\biggl(-\frac{\hbar\omega_{0}}{T} \biggl)\biggr] -{\cal N}_{0} \biggr].
\end{eqnarray}
Here $G_0 = (\overline{\Sigma_0}/\tau_0)\exp(-\hbar\omega_0/T_0)$ is the rate of the electron-hole pair generation due to the absorption of equilibrium  optical phonons  and
$\eta_{0} = \hbar^2\omega_{0}^2/\pi^2T^2$ characterizes the effect of the density of states in GLs on the interband and intraband optical phonon mediated transitions~\cite{58}. At $T_0 = 25$~meV, for the intra-valley optical phonons in GLs, $G_0 \simeq 10^{21}$~cm$^{-2}$s$^{-1}$~\cite{60}.

Considering the contribution of both the intra-valley and inter-valley optical phonons~\cite{60} as well as the interface optical phonons,
we set $G_0  \simeq 4\times 10^{21}$~cm$^{-2}$s$^{-1}$.
Setting $\hbar\omega_0 = 150$~meV, we obtain the
density of the hole states in GLs (involved in
the optical phonon absorption)
$\overline{\Sigma_0} = (2/\pi\hbar^2v_W^2)\int_0^{\hbar\omega_0}d\varepsilon\,\varepsilon = (1/\pi)(\hbar\omega_0/\hbar\,v_W)^2$,  
we obtain $\overline \Sigma_0\simeq 1.83\times 10^{12}$~cm$^{-2}$. Consequently, for  the optical phonon spontaneous emission: $\tau_0 = (\overline{\Sigma_0}/G_0)\exp(-\hbar\omega_0/T_0)\simeq 1.1$~ps. The ratio $\overline \Sigma_0/\Sigma_0$,
where $\Sigma_0 = (\pi/3)(T_0/\hbar\,v_W)^2$, which is the carrier density in the 2DEHP in equilibrium, is equal to $ 3(\hbar\omega_0/\pi\,T_0)^2 \simeq 10.9$.

 The Auger recombination-generation term in equation~(1) can be presented in the following simplified form (see, for example,~\cite{11}):

\begin{equation}\label{eqA3}
R_A \simeq \frac{{\overline{\Sigma_0}}}{\tau_A}\biggl[\exp\biggl(\frac{\mu_e + \mu_h}{T}\biggr) - 1\biggr].
\end{equation}

 The Auger mechanisms in GLs are very specific due to the linear gapless electron and hole dispersion law, which formally prohibits the recombination-generation processes with the participation of two electrons and one hole and two holes and one electron~\cite{65}. However, the Auger processes, 
involving other scatterers 
and affected by the dynamic screening of the carrier interactions and renormalization of the carrier energy spectra, can be essential. Moreover, the characteristic time of the Auger
recombination in GLs $\tau_A$,  being relatively long at moderate carrier densities and  temperatures,
can be fairly short for the high densities and temperatures. Indeed, as a thorough study shows~\cite{58}, 
$\tau_A \simeq 1.1$~ps at $T = 300$~K in the GL encapsulated by the hBNLs, but $\tau_A \simeq 0.07 - 0.2$~ps at $T = 1000 - 3000$~K. This implies that in the IDLEs, which can be effective just in the latter carrier  effective temperature range, we can assume (in contrast to some devices considered previously) that $\tau_A$ is smaller than $\tau_0$. 
The carrier-carrier scattering leads to the
effective carrier temperature common for the
electrons and holes. The Auger recombination-
generation processes tend to equilibrate the
electrons and holes aligning their quasi-Fermi
levels.
 Such an equilibrium (apart from the equality of the electron and hole effective temperatures) 
corresponds to $\mu_e = - \mu_h = \mu$, i.e., $\mu_e + \mu_h = 0$. In the case of undoped GLs, the latter
implies $\mu_e = \mu_h = \mu = 0$.

If $\tau_A$ is small, it follows from equation~(1) with equations~(A1) and (A2) that $2\mu/T = (\mu_e + \mu_h)/T$ is also small that leads to

\begin{eqnarray}\label{eqA4}
R_{0}^{inter} 
=\frac{\overline{\Sigma_0}}{\tau_0} 
\biggl[({\cal N}_{0} + 1)\exp\biggl(-\frac{\hbar\omega_{0}}{T} \biggl) -{\cal N}_{0} \biggr],
\end{eqnarray}

\begin{eqnarray}\label{eqA5}
R_{0}^{inter} +R_{0}^{intra}
= \frac{\overline{\Sigma_0}}{\tau_0} \biggl(1+\frac{1}{\eta_{O}}\biggr)\nonumber\\
\times 
\biggl[({\cal N}_{0} + 1)\exp\biggl(-\frac{\hbar\omega_{0}}{T} \biggl)
 -{\cal N}_{0} \biggr].
\end{eqnarray}
 Considering that, at the IDLE operation conditions the Auger recombination time $\tau_A$ is fairly small  
 and
expressing the quantity $\exp[(\mu_e +\mu_h)/T]$ via $i_0$, from  equation (6) we obtain

\begin{eqnarray}\label{eqA6} 
\biggl(1 +\frac{1}{\eta_0}\biggr) \biggl[({\cal N}_{0} + 1)\exp\biggl(-\frac{\hbar\omega_{0}}{T} \biggl) -{\cal N}_{0} \biggr]\nonumber\\
 = i_0\biggl[\frac{\overline{\Delta_i}}{\hbar\omega_0}
-  ({\cal N}_0 + 1)\exp\biggl( -\frac{\hbar\omega_0}{T}\biggr)\frac{\tau_A}{\tau_0}\biggr]. 
\end{eqnarray}

\subsection*{2. Carriers thermal energy and their heat capacitance}

 The density of the carrier energy in the 2DEHP can be calculated as

\begin{eqnarray}\label{eqA7}
{\cal E} =\frac{2T^3}{\pi\hbar^2v_W^2}\int_0^{\infty}dzz^2\nonumber\\
\times\biggl[\frac{1}{1 +\exp\biggl(z - \mu_e/T\biggr)}
+ \frac{1}{1+\exp\biggl(z - \mu_h/T\biggr)}\biggr]\nonumber\\
%\simeq\frac{\pi\,T^2}{6\hbar^2v_W^2} (2T + \mu_e + \mu_h)\nonumber\\
%\simeq 
%\frac{2T^3}{\pi\hbar^2v_W^2}\int_0^{\infty}dz\biggl[\frac{2 z^2}{(1 + e^z)}
%+ \frac{\mu_e+\mu_h}{T}\frac{z^2e^z}{(1 + e^z)^2)}\biggr]
%\nonumber\\
\simeq \frac{2T^3}{\pi\hbar^2v_W^2}\biggl(3\zeta(3) + 1.64\frac{\mu_e+\mu_h}{T}\biggr)\nonumber\\
%\simeq \frac{2T^3}{\pi\hbar^2v_W^2}\biggl(3.61 + 1.64\frac{\mu_e+\mu_h}{T}\biggr)\nonumber\\
\simeq \nu\Sigma_0 T \biggl(\frac{T}{T_0}\biggr)^2,
%\times\biggl[\frac{1}{1 +\exp\biggl(z - \mu_e/T\biggr)}
%\frac{2\pi^2\overline{\Sigma_0}}{3} \biggl(\frac{T}{\hbar\,\omega_0}\biggr)^2T.
\end{eqnarray}
where 
$\nu = 18\zeta(3)/\pi^2 \simeq 6\cdot 3.61/\pi^2 \simeq 2.19$ and 
$\zeta(x)$ is the Rieman zeta-function.

At $|\mu_e + \mu_h| \ll T$, equation~(A7) yields the following value of the 2DEHP specific heat capacitance per unit of the 2DEHP area:

\begin{eqnarray}\label{eqA8}
C_c = 3\nu \Sigma_0\biggl(\frac{T}{T_0}\biggr)^2 = \pi\nu\biggl(\frac{T}{\hbar\,v_W}\biggr)^2,
\end{eqnarray}
where
$c_c = 3\nu \simeq 6.57 $ stands for the carrier heat capacitance per one carrier.

The carrier specific heat capacitance per unit area $C_c =c_c\Sigma_0(T/T_0)^2$ is small compared with the specific lattice heats of the GLs and hBNLs. However,  $C_c$ determines
the high-speed modulation characteristic (the maximum modulation frequency $f_m^{max} \propto C_c$) since, in this case, the lattice temperature does not  follow the external modulation signals.

\subsection*{3. Distribution of the injected energy between the 2DEHP and the optical phonon system}

Considering the cascade of optical phonons emitted by the injected holes, on can estimate
the fractions of the power, $\hbar\omega_0k_0/\Delta_i$, which directly transferred to the optical phonon systems, invoking the following formula:

\begin{eqnarray}\label{eqA9}
k_{0} = \sum_{n =1}^{K_{0}} \frac{1}{(1 + \tau_0/\tau_{cc})^n}\nonumber\\
 = \frac{1}{\tau_0/\tau_{cc}}
\biggl[1 -  \frac{1}{(1 + \tau_0/\tau_{cc})^{K_{0}}}\biggr]
\end{eqnarray}
with $K_{0}$ being the number of the pertinent optical phonons in the cascades.
In the case of small  $\tau_0/\tau_{cc}$, equation~(A9) can presented as $k_{0} \simeq K_{0}/(1 + K_{0}\tau_0/\tau_{cc})$.

Equation~(A9) ignores the dependence of $\tau_0$ on the injected hot hole energy. In the case when $\tau_0$ is proportional to the hole energy, one can obtain the following formula replacing equation~(A9):

\begin{eqnarray}\label{eqA10}
k_{0} = \frac{1}{[1 + \tau_0/K_{0}\tau_{cc}]}\nonumber\\
 + 
\frac{1}{[1 + \tau/K_{0}\tau_{cc}] [1 + \tau_0/(K_{0}-1)]} +....\nonumber\\ 
= \frac{K_{0}}{1 + \tau_0/\tau_{cc}}.
\end{eqnarray}

\end{document}